\documentclass[aps,superscriptaddress,notitlepage,reprint]{revtex4-1}

\usepackage{epsfig,color}
\usepackage{graphicx}
\usepackage{dcolumn}
\usepackage{bm}
\usepackage{amsmath, amsfonts, amssymb,mathrsfs}
\usepackage{pstricks}
\usepackage{amsxtra}
\usepackage{cases}
\usepackage{amsthm}
\usepackage{braket}
\usepackage{natbib}
\usepackage{physics}
\usepackage{hyperref,color}
\usepackage{multirow}
\definecolor{darkblue}{rgb}{0.0,0.0,0.7}
\hypersetup{colorlinks,breaklinks,linkcolor=darkblue,urlcolor=darkblue,anchorcolor=darkblue,citecolor=darkblue}

\providecommand{\ad}{{\hat{a}^\dagger}}

\providecommand{\bd}{{\hat{b}^\dagger}}
\providecommand{\mh}[1]{\hat{m}_{#1}} 
\providecommand{\md}[1]{\hat{m}_{#1}^{\dagger}} 
\providecommand{\Hh}{{\hat{H}}}
\providecommand{\ka}[1]{\kappa_{#1}}
\providecommand{\g}[1]{g_{#1}}
\providecommand{\D}[1]{\Delta_{#1}}
\providecommand{\ch}[1]{\chi_{#1}}

\begin{document}
\title{Generalized matching condition for unity efficiency quantum transduction}

\author{Chiao-Hsuan Wang}
\email{chiaowang@phys.ntu.edu.tw}
\affiliation{Department of Physics and Center for Theoretical Physics, National Taiwan University, Taipei 10617, Taiwan}
\affiliation{Center for Quantum Science and Engineering, National Taiwan University, Taipei 10617, Taiwan}
\affiliation{Physics Division, National Center for Theoretical Sciences, Taipei, 10617, Taiwan}
\affiliation{Pritzker School of Molecular Engineering, University of Chicago, Chicago, Illinois 60637, USA}

\author{Mengzhen Zhang}
\affiliation{Pritzker School of Molecular Engineering, University of Chicago, Chicago, Illinois 60637, USA}

\author{Liang Jiang}
\affiliation{Pritzker School of Molecular Engineering, University of Chicago, Chicago, Illinois 60637, USA}

\begin{abstract}
Coherently converting quantum states between distinct elements via quantum transducers remains a crucial yet challenging task in quantum science.  Especially in demand is quantum transduction between optical frequencies, which are ideal for low-loss transmission across long distances, and microwave frequencies, which admit high-fidelity quantum operations.  We present a generic formalism for $N$-stage quantum transduction that covers various leading microwave-to-optical, microwave-to-microwave, and optical-to-optical linear conversion approaches.  We then identify effective circuit models and the resulting generalized matching conditions for achieving maximum conversion efficiency.  The generalized matching condition requires resistance matching as well as frequency matching beyond the usual resonant assumption, with a simple impedance-matched transmission interpretation.  Our formalism provides a generic toolbox for determining experimental parameters to realize efficient quantum transduction and suggests different regimes of non-resonant conversions that might outperform all-resonant ones.
\end{abstract}
\maketitle
 
Global quantum networks for secured communication,  distributed quantum computation, and beyond is among the appealing applications of future quantum technologies~\cite{Elliott2002,Kimble2008,Komar2014,Simon2017}.
Lying at the heart of networking quantum devices are quantum transducers, coherent interfaces that can faithfully transmit quantum information between components with distinct energies ~\cite{Lauk2020,Lambert2020,Han2021}.  Major interests have been placed on coherent conversions between optical and microwave frequencies, with the former ideal for reliable long-range communications through optical fibers or in free space~\cite{Takesue2015,Yin2017}, and the latter admitting high-fidelity local quantum operations using superconducting processors~\cite{Blais2021,Joshi2021}.

A variety of physical implementations for direct state transfer between microwave and optical systems have been rapidly developed in the past decade. Direct conversion between microwave and optical frequencies can be established through electro-optics~\cite{Tsang2010,Javerzac2016,Rueda2016,Fan2018,Fu2021}, electro-optomechanics~\cite{Andrews2014,Higginbotham2018}, opto-magnonics~\cite{Hisatomi2016,Zhu2020}, piezo-optomechanics~\cite{Han2020b,Mirhosseini2020}, and atom-assisted conversion schemes that involve two or more intermediate atomic levels~\cite{Han2018,Everts2019,Bartholomew2020,Tu2022,Asadi2022}. Despite the wide range of experimental settings, these approaches can be modeled as multistage coupled bosonic chains with two end modes coupled externally.  The same model applies to optomechanical photon-phonon translators~\cite{Safavi-Naeini2011}, microwave-to-microwave frequency converter~\cite{Abdo2013} in superconducting circuits, and optical-to-optical frequency converters via cavity-optomechanics~\cite{Hill2012} or coupled-resonator optical waveguides through eight-ring resonators~\cite{Morichetti2011}.

When the environmental thermal noise is negligible or can be fully suppressed by cooling \cite{Lecocq2016,Xu2020}, the performance of a direct quantum transducer is measured by the transmissivity $\eta$, i.e., the coherent conversion efficiency between the input signal and the output signal~\cite{Zeuthen2020}. A perfect direct quantum transducer should be of unity efficiency and hence can faithfully transfer quantum signals with zero information loss. With more complicated experimental implementations, such as three- to six-wave mixing atom-based transducers~\cite{Han2018,Everts2019,Bartholomew2020}, being pursued, a systematic approach to designing physical parameters for achieving unity efficiency transduction is pressing.

In this Letter, we derive a unity-efficiency-achieving condition inspired by the widely known impedance-matching condition in electrical engineering. Based on this discovery, we establish an effective electric circuit model to clarify the analogy between direct quantum transduction and electrical power transmission. The generality of our unity-efficiency-achieving condition allows us to loosen the stringent constraints imposed on previous quantum transduction schemes, such as the requirement that all involved modes must be on resonance in the rotating frame to achieve maximum efficiency (after consideration of frequency shifts given by the rotating-wave approximation) ~\cite{Safavi-Naeini2011,Andrews2014,Williamson2014}, which may extend the applicability and flexibility of direct quantum transduction. We also investigate the robustness of this condition and its working range under the influence of environmental dissipation introduced to the intermediate stages.

\begin{figure}[htbp]
\begin{center}
\includegraphics[width=0.5 \textwidth]{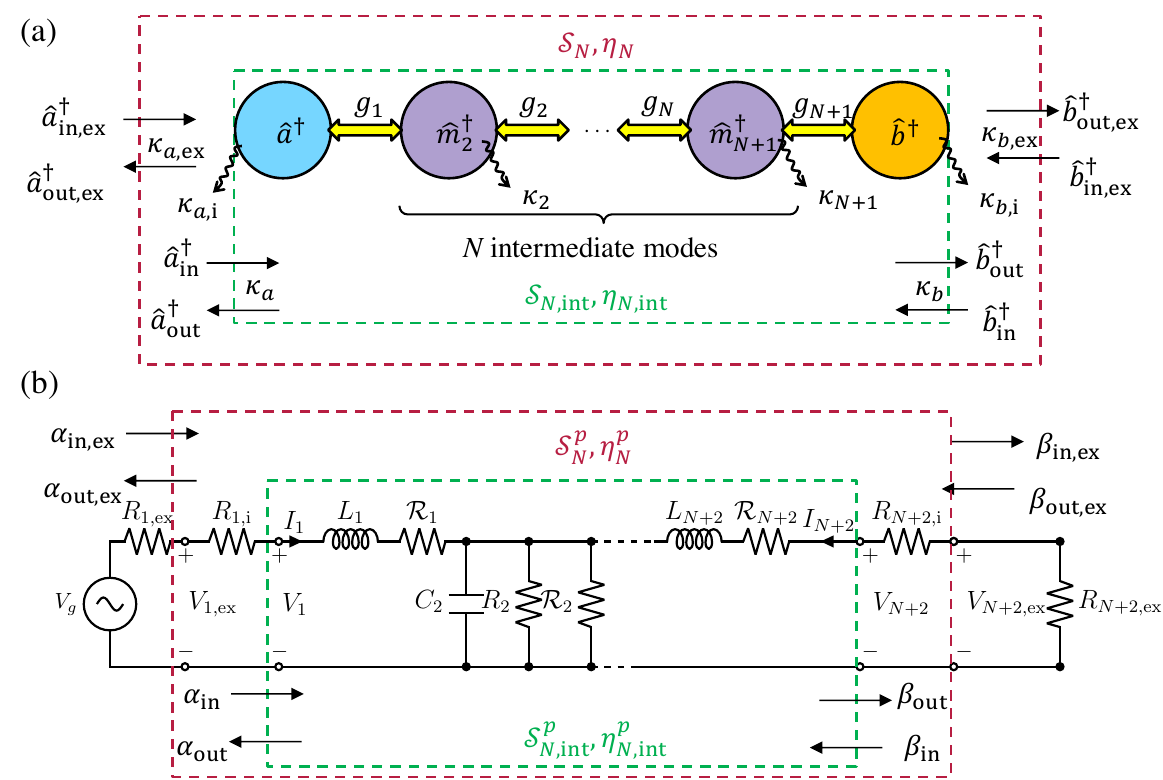}
\caption{(a) Schematic of the $N$-stage quantum transduction system with photon number conversion efficiency $\eta_N$ and internal photon number conversion efficiency $\eta_{N,\rm int}$. (b) Schematic of an effective circuit network for $N$-stage transduction (with odd $N$) with power efficiency $\eta^p_N=\eta_N$ and internal power efficiency $\eta^p_{N,\rm int}=\eta_{N,\rm int}$.}
\label{fig:nstage} 
\end{center}
\end{figure}

\paragraph*{Model of $N$-stage direct quantum transduction.}
The transduction device we are considering is a bosonic chain consisting of $N+2$ modes, where each mode in this chain is coupled to their nearest neighbors through passive interactions [Fig.~\ref{fig:nstage}(a)] regardless of their physical carriers (e.g. optical modes, microwave modes, acoustic modes, magnons, etc.). The implementation of such systems has been demonstrated in existing quantum transduction experiments~\cite{Tsang2010,Javerzac2016,Rueda2016,Fan2018,Fu2021,Andrews2014,Higginbotham2018,Hisatomi2016,Zhu2020,Han2020b,Mirhosseini2020,Han2018,Everts2019,Bartholomew2020,Tu2022,Asadi2022,Safavi-Naeini2011,Abdo2013,Hill2012,Morichetti2011} by properly driving the involved modes.  The whole device can be described by the following Hamiltonian
\begin{align}
\Hh_N=-\sum_{\rm j=1}^{N+2}  \D{j} \md{j} \mh{j} + \sum_{\rm j=1}^{N+1}\g{j}\left(\md{j}\mh{j+1}+ \md{j+1}\mh{j}\right),
\label{eqn:HN}
\end{align}
with $\hat{m}_j$, $\hat{m}^\dagger_j$ the annihilation and creation operators of each mode, $\Delta_j$ the detuning of each mode in the rotating frame, and $g_j$ the coupling strength of each interaction. To stress the presence of the intermediate modes, we address each of them as a \textit{stage}. Therefore, a device with $N+2$ modes contains $N$ stages as the intermediate modes. For similar reasons, we use $\hat{a}$ as an alternative denotation for $\hat{m}_1$ and $\hat{b}$ for $\hat{m}_{N+2}$ as shown in Fig.~\ref{fig:nstage}(a).

For transduction, this device will be passively coupled with external channels to receive the input signal $\hat{a}^{\dagger}_{\rm in, ex}$ at the rate $\kappa_{a, \rm ex}$ and to emit the output signal $\hat{b}^{\dagger}_{\rm out, ex}$ at the rate $\kappa_{b, \rm ex}$, requiring the modes $\hat{a}^{\dagger}$ and $\hat{b}^{\dagger}$ to be of disparate frequencies (e.g., $\hat{a}^{\dagger}$ may be a microwave-mode while is $\hat{b}^{\dagger}$ an optical mode).  Taking environmental dissipation into consideration, we introduce an intrinsic loss rate $\kappa_j$ for each mode. The intrinsic loss rates for the first and the last modes are referred to as $\kappa_{a, \rm i}$ and $\kappa_{b, \rm i}$, respectively, to be consistent with our conventions. Thus, the rates $\kappa_a = \kappa_{a, \rm i} + \kappa_{a, \rm ex}$ and $\kappa_b = \kappa_{b, \rm i} + \kappa_{b, \rm ex}$ are used to denote the total dissipation rates of the two modes. Similarly, the other parameters, $\Delta_1$, $\Delta_{N+2}$, $g_1$, $g_{N+1}$ are referred to as $\Delta_a$, $\Delta_b$, $g_a$, $g_b$, respectively.

The conversion efficiency $\eta$ of the transducer quantifies the fraction of photon (boson) number power that can be successfully transferred from one end to the other. Treating the $N$-stage transducer as a scattering matrix $\mathcal{S}_N$ between the external input and output modes as shown by the red dashed boundary in Fig.~\ref{fig:nstage}(a), the photon number conversion efficiency from $\ad$ to $\bd$ is defined as $\eta_N \equiv \abs{\mathcal{S}_{N\bd_{\rm out, ex}\ad_{\rm in, ex}}}^2$. Here we have been focused on the scheme of a bosonic chain with only nearest-neighbor beam-splitter interactions such that there is neither nonreciprocity ~\cite{Ranzani2015} nor amplification effects~\cite{Clerk2010}. A perfect transduction is well-defined in this scenario by the unity efficiency condition $\eta_N=1$.

The $N$-stage conversion efficiency at a signal frequency $\omega$ can be expressed in a general form~\footnote{See Supplemental Information for detailed derivations and discussions,
explicit expressions for the parameter regimes,
optimal frequencies, and maximal achievable internal efficiencies,
and expressions of added noise when the system satisfies
the generalized matching condition.}
\begin{align}
\eta_N[\omega]=\frac{\kappa_{a, \rm ex} \kappa_{b, \rm ex}}{\kappa_{a} \kappa_{b}} \abs{\frac{\sqrt{\kappa_{a} \kappa_{b}}\prod_{ j=1}^{N+1}i\g{j}}{\prod_{ j=1}^{N+2}\ch{j,\rm eff}^{-1}}}^2=\eta_{\rm ext} \eta_{N,\rm int}[\omega],
\label{eqn:etaN}
\end{align}
where $\ch{j,\rm eff}$ is the effective susceptibility for $\hat{m}^{\dagger}_j$, a modification to the bare mode susceptibility $\ch{j}\equiv [i (\omega+\D{j})+\kappa_j/2]^{-1}$, due to the couplings $\g{j},\cdots, g_{N+1}(=g_b)$,
\begin{align}
\ch{j,\rm eff}^{-1} \equiv \ch{j}^{-1}+\cfrac{\g{j}^2}{\ch{j+1}^{-1}+\cfrac{\ddots}{\ddots+\cfrac{\g{b}^2}{\ch{b}^{-1}}}}.
\label{eqn:chieff}
\end{align}
Here, $\eta_{\rm ext}=\kappa_{a, \rm ex} \kappa_{b, \rm ex}/\kappa_{a} \kappa_{b}$ is the external conversion efficiency and $\eta_{N,\rm int}[\omega]$ is the $N$-stage internal conversion efficiency.

The internal efficiency $\eta_{N,\rm int}[\omega]$ can be understood as the efficiency of the internal scattering process $\mathcal{S}_{N,\rm int}$ as shown by the green dashed boundary in Fig.~\ref{fig:nstage}(a). The relevant scattering modes are the total input and output modes $\ad(\bd)_{\rm in/out}$ coupled to $\ad(\bd)$ at an overall rate $\kappa_{a(b)}$, due to the interaction with both the external channels and the dissipative environment.

The external efficiency $\eta_{\rm ext}$, on the other hand, represents the average fraction of information that can be successfully transferred through the external channels, without being lost to the dissipative environments at rates $\kappa_{a,\rm i}$ and $\kappa_{b,\rm i}$, while leaving or entering the end modes $\ad$ and $\bd$. In the strong external coupling limit such that $\kappa_{a, \rm ex} \gg \kappa_{a, \rm i}$ and $\kappa_{b, \rm ex} \gg \kappa_{b, \rm i}$, one can reach maximal external efficiency $\eta_{\rm ext}=1$.  We will thus focus on the general criteria for system parameters to attain unity internal conversion efficiency.   Since the system dynamics is invariant under an global energy shift, we will use $\nu_{j}\equiv\omega+\D{j}$ as the relevant frequency variables.

\begin{figure}[htbp]
\begin{center}
\includegraphics[width=0.48 \textwidth]{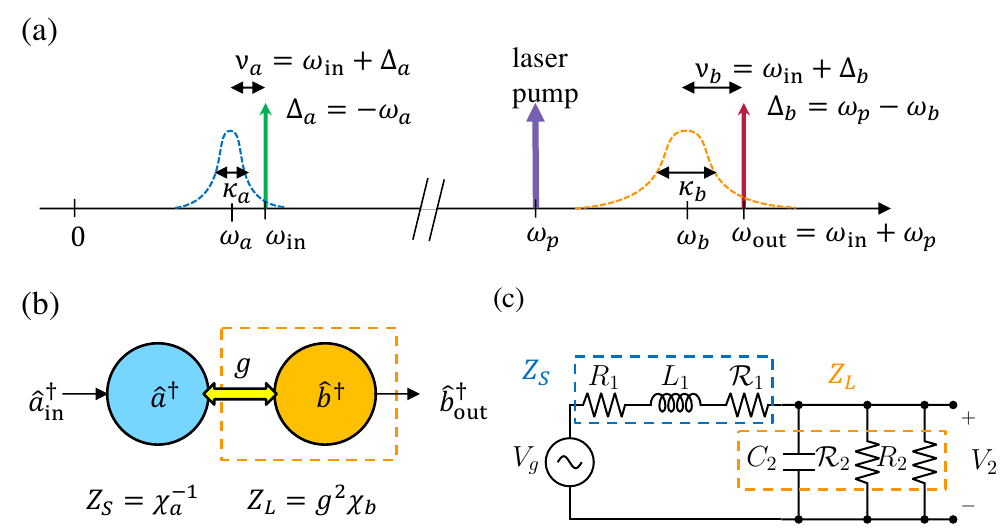}
\caption{(a) An exemplar frequency diagram for 0-stage transduction from $\ad$ to $\bd$ in the laboratory frame. Here, $\omega_{a(b)}$ is the frequency of mode $\ad(\bd)$ and $\omega_{\rm in(out)}$ is the frequency of the input (output) signal. The system can be described with a time-independent Hamiltonian as in Eq.~(\ref{eqn:HN}) by moving $\bd$ and $\bd_{\rm out,ex}$ to the rotating frame at the laser pump frequency $\omega_p$ that mediates the linear up- and down- conversion. (b) Schematic of the 0-stage quantum transduction model. (c) Effective circuit model that reproduces the 0-stage internal conversion efficiency by setting $L_1=1$, $\mathcal{R}_1=i \D{a}$, $R_1=\kappa_{a}/2$, $C_2=1/g^{2}$, $\mathcal{R}_2=g^2/(i\D{b})$, and $R_2=2g^2/\kappa_{b}$.}
\label{fig:0stage} 
\end{center}
\end{figure}

As a starting example, we study the simplest case of $0$-stage transduction [Fig.~\ref{fig:0stage}(b)] relevant for electro-optical quantum transducers~\cite{Tsang2010,Javerzac2016,Rueda2016,Fan2018,Fu2021}.  The $0$-stage internal efficiency is
\begin{align}
\eta_{0,\rm int}=\abs{\frac{g \sqrt{\kappa_{b}\kappa_{a}}}{g^2+\ch{a}^{-1}\ch{b}^{-1}}}^2.
\label{eqn:eta0}
\end{align}
Under the traditional all-resonant assumption $\nu_a(\equiv\omega+\D{a})=\nu_b(\equiv\omega+\D{b})=0$, the matching condition to fulfill $\eta_{0,\rm int}=1$ was given by $\mathcal{C}_{a,b}\equiv4 g^2/\kappa_{a}\kappa_{b}=1$, where $\mathcal{C}_{a,b}$ is the cooperativity between modes $\ad$ and $\bd$~\cite{Safavi-Naeini2011}.  On the other hand, two off-resonant peaks in efficiency have also been observed in the strong-coupling regime $\mathcal{C}_{a,b}>1$ with $\kappa_{a}=\kappa_{b}$~\cite{Tsang2011}. However, there is no systematic analysis of these off-resonant peaks.

A natural question arises: What is the most general condition to achieve unity internal conversion efficiency?
Solving for the solution of $\nu_{a}\equiv \omega+\D{a}$ given $\eta_{0,\rm int}[\omega]=1$, we arrive at a set of independent criteria
\begin{numcases}{}
\nu_{a}=\frac{g^2 \nu_{b}}{\nu_{b}^2+\kappa_{b}^2/4}, \label{eqn:nua}\\
\kappa_{a}=\frac{g^2 \kappa_{b}}{\nu_{b}^2+\kappa_{b}^2/4} \label{eqn:ka},
\end{numcases}
where $\nu_{b}\equiv \omega+\D{b}$.
The all-resonant condition $\nu_{a}=\nu_{b}=0$ fulfills the nonlinear condition of Eq.~(\ref{eqn:nua}), which might have other solutions. Before proceeding with the general solution to the matching condition, we would like to point out that the conditions of Eqs.~(\ref{eqn:nua}) and~(\ref{eqn:ka}) admit an impedance-matching interpretation analogous to the electric circuit power transfer theory.

\paragraph*{Impedance matching interpretation for unity efficiency transduction.}
 In electric circuits, for an active voltage transferred from a source impedance $Z_S$ to a load impedance $Z_L$, maximum transfer power is obtained if the conjugate impedance matching condition $Z_L=Z_S^{*}$ is satisfied~\cite{Rahola2008}.
For the 0-stage transduction system, unity internal conversion efficiency is also achieved when the system has impedance matched parameters.  Viewed from the $\ad$ port, $\chi^{-1}_{a,\rm eff}$ is modified by $g^2 \ch{b}$ due to the coupling with the $\bd$ mode.  By interpreting the original inverse susceptibility of mode $\ad$ as the source impedance in the unit of rates, $Z_S=\ch{a}^{-1}$, and the couplng-induced inverse susceptibility modification as the load impedance, $Z_L=g^2 \ch{b}$, the first condition Eq.(\ref{eqn:nua}) is equivalent to having $\text{Im}\ch{a}^{-1}=-\text{Im}g^2 \ch{b}$ while the second condition Eq.(\ref{eqn:ka}) corresponds to $\text{Re}\ch{a}^{-1}=\text{Re}g^2 \ch{b}$, which altogether fulfill $Z_L=Z_S^{*}$.

\begin{table}
\caption{Optimal frequencies for 0-stage transduction}
\begin{ruledtabular}
\begin{tabular}{ l l l }
$N$=0  & $ \mathcal{C}_{a,b} > 1 $& $\mathcal{C}_{a,b} \leq 1$ \\
 \hline
$\nu_{a}\equiv\omega+\D{a}$  & $\pm \frac{\kappa_{a}}{2}\sqrt{\mathcal{C}_{a,b}-1}$& 0 \\
$\nu_{b}\equiv\omega+\D{b}$ & $\pm \frac{\kappa_{b}}{2}\sqrt{\mathcal{C}_{a,b}-1}$ & 0\\
$\eta_{\rm 0,int}^{\rm max}$ & 1 & $\frac{4 \mathcal{C}_{a,b}}{(\mathcal{C}_{a,b}+1)^2}$
\end{tabular}
\end{ruledtabular}
\label{table:0stage}
\end{table}

To construct a rigorous connection between maximal conversion efficiency for transducers and the impedance-matching condition for electric circuits, we now establish an effective circuit model. Specifically, the effective circuit shown in Fig.~\ref{fig:nstage}(b) can reproduce an identical frequency response as a $N$-stage quantum transducer in Fig.~\ref{fig:nstage}(a) with an odd $N$, such that the electric power internal efficiency is the same as the photon number internal efficiency $\eta^{p}_{N\rm ,int}[\omega]=\eta_{N\rm ,int}[\omega]$, by setting~\cite{Note1}
\begin{align}
\begin{cases}
L^{-1}_{j} C^{-1}_{\rm j+1}=\g{j}^{2}, \,R_{j} L^{-1}_{j} =\ka{j}/2, \, \mathcal{R}_{j} L^{-1}_{j} = i \Delta_j  & \text{odd }j,\\
C^{-1}_{j} L^{-1}_{\rm j+1}=\g{j}^{2}, \, G_{j} C^{-1}_{j} =\ka{j}/2, \, \mathcal{G}_{j} C^{-1}_{j} = i \D{j} & \text{even }j.
\end{cases}
\label{LC}
\end{align}
The circuit is composed of elements with inductance $L_{j} \in \mathbb{R}$, capacitance $C_{j} \in \mathbb{R}$, resistance $R_{j}=1/G_{j} \in \mathbb{R}$, and generalized resistance $\mathcal{R}_{j}=1/\mathcal{G}_{j} \in i \mathbb{R}$.  The resistance at the two ends can be separated into an external coupling and intrinsic loss components, $R_{1(N+2),\rm ex} L^{-1}_{1(N+2)}=\ka{a(b), \rm ex}/2$ and  $R_{1(N+2),\rm i} L^{-1}_{1(N+2)}=\ka{a(b),\rm i}/2$, to replicate the total conversion efficiency $\eta^{p}_{N}[\omega]=\eta_{N}[\omega]$. Here, we have introduced the \textit{generalized} resistance of imaginary values $\mathcal{R}_{j}$ to account for the independent mode detunings $\Delta_j$, which is distinct from prior numerical~\cite{Yariv2011} or synthetic methods~\cite{Van2006,Naaman2021} for coupled resonator arrays.

With the explicit circuit correspondence, we see a clear analogy between the scattering process of the input/output modes, $\ad(\bd)_{\rm in/out}$, for transducers and the incident/reflective power waves, $\alpha(\beta)_{\rm in/out}$, for electric circuits~\cite{Youla1971,Kurokawa1965,Rahola2008}. With the effective circuit parameters shown in Fig.~\ref{fig:0stage}(c), the impedance-matching analogy for attaining unity 0-stage internal efficiency naturally follows through.

\paragraph*{Generalized matching condition.} We now extend the concept of impedance matching to $N$-stage quantum transduction. An impedance-matching condition viewed from the $\ad$ mode (taking $L_1=1$) reads
\begin{align}
(\ch{a}^{-1})^{*}=\g{a}^2 \chi_{2,\rm eff}.
\label{eqn:amatch}
\end{align}
When all the intermediate modes are lossless, $\ka{2}= \cdots=\kappa_{N+1}= 0$, this condition also leads to impedance matching viewed from the $\bd$ mode (taking $L_{N+2}$ or $C_{N+2}=1$),
\begin{align}
(\ch{b}^{-1})^{*}=\g{b}^2 \chi_{N+1,\rm eff,r},
\label{eqn:bmatch}
\end{align}
where $\ch{j,\rm eff,r}$ is the effective susceptibility of mode $j$ viewed from the reversed direction due to the couplings $g_{1}(=g_a), \cdots ,g_{j-1}$,
\begin{align}
\ch{j,\rm eff,r}^{-1} \equiv \ch{j}^{-1}+\cfrac{\g{j-1}^2}{\ch{j-1}^{-1}+\cfrac{\ddots}{\ddots+\cfrac{\g{a}^2}{\ch{a}^{-1}}}}.
\label{eqn:chieffr}
\end{align}
One can achieve $\eta_{N, \rm int}=1$ by requiring lossless condition $\ka{2}= \cdots=\kappa_{N+1}= 0$ together with an impedance matching condition Eqs.~(\ref{eqn:amatch}) and ~(\ref{eqn:bmatch}).

There exist additional physical interpretations for the above matching conditions Eq.~(\ref{eqn:amatch}\&\ref{eqn:bmatch}).  First, they give rise to zero internal reflections. Similar to the fact that the power wave reflection coefficient at the source, $r^{p}_{S}=(Z_L-Z_S^*)/(Z_L+Z_S)$, vanishes for a conjugate matched lossless electric circuit, the above conditions also suggest zero internal reflection for the transducer. Specifically, the internal reflection coefficients are given by
\begin{align}
&r_{N,a}=\mathcal{S}_{N,\textrm{int } \ad_{\rm out}\ad_{\rm in}}=\frac{\g{a}^2 \chi_{2,\rm eff}-(\ch{a}^{-1})^{*}}{\g{a}^2 \chi_{2,\rm eff}+\ch{a}^{-1}},\\
&r_{N,b}=\mathcal{S}_{N,\textrm{int } \bd_{\rm out}\bd_{\rm in}}=\frac{\g{b}^2 \chi_{N+1,\rm eff,r}-(\ch{b}^{-1})^{*}}{\g{b}^2 \chi_{N+1,\rm eff,r}+\ch{b}^{-1}}.
\label{eqn:reflectionless}
\end{align}
It is clear that Eq.~(\ref{eqn:amatch}) implies $r_{N,a}=0$ while Eq.~(\ref{eqn:bmatch}) implies $r_{N,b}=0$.
Second, the matching conditions also correspond to critical effective cooperativities along the transduction chain when the system is lossless,
\begin{align}
\mathcal{C}_{j,j+1}^{\rm eff}\equiv \frac{g_j^2}{(\ch{j,\rm eff,r}^{-1})^{*}\chi_{j+1,\rm eff}^{-1}}=1.
\label{eqn:Ceff}
\end{align}

We summarize the generalized $N$-stage matching condition in a matrix determinant form~\cite{Note1},
\begin{align}
M_N=0,
\label{eqn:MatchingN}
\end{align}
\begin{align}
M_{N}\equiv \left.\begin{vmatrix}
-(\ch{a}^{-1})^{*} & i\g{a} & 0& \cdots & \cdots &0\\
 i\g{a} & \ch{2}^{-1} & i g_2  &\ddots & & \vdots\\
0 & i\g{2} & \ddots & \ddots  & \ddots  &  \vdots\\
\vdots & \ddots &  \ddots   & \ddots  & \ddots &  0 \\
\vdots &  & \ddots  & \ddots & \ddots &  i g_{b}\\
 0 & \cdots & \cdots  & 0 & i g_{b}&\ch{b}^{-1}
\end{vmatrix} \right|_{\substack{\kappa_{2}=\cdots\\=\kappa_{N+1}=0}}.
\label{eqn:MN}
\end{align}
We call the real part of the equation the generalized resistance matching condition, $\text{Re}M_{N}=0$ and the imaginary part of the equation the generalized resonant condition, $\text{Im}M_{N}=0$.

The generalized matching condition Eq.~(\ref{eqn:MatchingN}) takes a symmetric form in $a$ and $b$. For example, the generalized 0-stage matching condition reads
\begin{numcases}{M_0=0\Rightarrow \\}
\text{Re} : g^2=\frac{\kappa_{a}\kappa_{b}}{4}+\nu_{a}\nu_{b}, \label{MCR0}\\
\text{Im} : \frac{\nu_{a}}{\kappa_{a}}=\frac{\nu_{b}}{\kappa_{b}}, \label{MCI0}
\end{numcases}
which is mathematically equivalent to the previous set of criteria Eqs.~(\ref{eqn:nua}) and ~(\ref{eqn:ka}) for getting $\eta_{0,\rm int}[\omega]=1$.

Note that in quantum transduction systems there exist $\D{i}$ degrees of freedom associated with the generalized resistance of imaginary values, which are unavailable in real-world linear electric circuits where the resistors are always real. The above expression is thus a generalized version of a matching condition beyond the all-resonant assumption $\forall j, \nu_{j} \equiv \omega + \D{j}=0$, which corresponds to circuits with purely real resistors while setting the signal frequency to be zero, $\omega=0$, in the rotating frame.

\paragraph*{Tuning the off-resonant frequencies.}
In a typical analysis, the frequencies $\nu_j${'}s, after including the frequency shifts given by the rotating-wave approximation, if any, are often chosen to be zero (all resonant) in order to achieve maximum conversion efficiency~\cite{Safavi-Naeini2011,Andrews2014,Williamson2014}.  Here, we consider generic schemes such that $\nu_j${'}s may be nonzero (off-resonant) to offer extra tunability for transduction. One might adjust the system parameters to achieve an ideal conversion for output signal frequencies detuned from $\omega_b$. In the strong-coupling regime $\mathcal{C}_{a,b}>1$ of 0-stage transduction, unity internal efficiency is achieved with non-resonant frequencies $\nu_a \neq 0$ and $\nu_b \neq 0$. The optimal frequencies and the maximal internal efficiency for a given $\mathcal{C}_{a,b}$ are summarized in Table~\ref{table:0stage}. Note that the transduction process still follows energy conservation even with non-resonant $\nu_{j}$'s. For example, $\omega_{\rm out}=\omega_{\rm in}+\omega_{p}$ is always true for the 0-stage case shown in Fig.~\ref{fig:0stage}(a), while $\nu_{j}$'s merely represent the frequency differences between the input/output signals and modes $\ad$ or $\bd$. 
 
The extra degrees of freedom given by non-resonant $\nu_j${'}s can also serve as optimizable parameters in experiments.  In practice, the loss rate of the intermediate modes may be non-negligible, and the system can no longer satisfy $M_N=0$ to reach unity internal efficiency.  The mode cooperativities $\mathcal{C}_{j,j+1}\equiv 4g_j^2/(\kappa_j \kappa_{j+1})$ are thus finite and limited by experimental constraints.  On the other hand, the frequencies and detunings of the modes are typically tunable by laser drives. In such experimental settings, one can find the optimal $\nu_j${'}s that lead to maximal efficiency at the given values of $\mathcal{C}_{j,j+1}${'}s.

We find that the optimal parameters with intermediate losses can again be interpreted as impedance-matched parameters of lossy circuits.
Take the case of 1-stage transduction, appropriate for  electro-optomechanical~\cite{Andrews2014,Higginbotham2018}, opto-magnonical~\cite{Hisatomi2016,Zhu2020}, or piezo-optomechanical~\cite{Han2020b,Mirhosseini2020} quantum transducers, for example.
When $\kappa_2 \neq 0$, we can define mode cooperativities $\mathcal{C}_{a,2}=\g{a}^2/\kappa_{a}\ka{2}$ and $\mathcal{C}_{2,b}=\g{b}^2/\kappa_{2}\ka{b}$. When one of the modeS is overcoupled, for instance, if $\mathcal{C}_{a,2} > \mathcal{C}_{2,b}+1$, a relevant regime for piezo-optomechanical transducers in which the microwave mode is overcoupled ~\cite{Han2020b,Mirhosseini2020}, the maximal internal efficiency is achieved by the choice of optimal frequencies satisfying $ (\ch{a}^{-1})^{*}=\g{a}^2/(\ch{2}^{-1}+\g{b}^2 \ch{b})$ and $\nu_b=0$. This corresponds to impedance matching at the $\ad$ mode while treating $\md{2}$ and $\bd$ altogether as the load. If $\abs{\mathcal{C}_{a2} - \mathcal{C}_{2b}}\leq 1$, the conversion is optimized when the system is all resonant.
This condition can again be understood as an impedance matched lossy circuit while treating the middle mode as a lossy component partly in the source and partly in the load. ~\cite{Note1}.

\begin{figure}[htbp]
\begin{center}
\includegraphics[width=0.5 \textwidth]{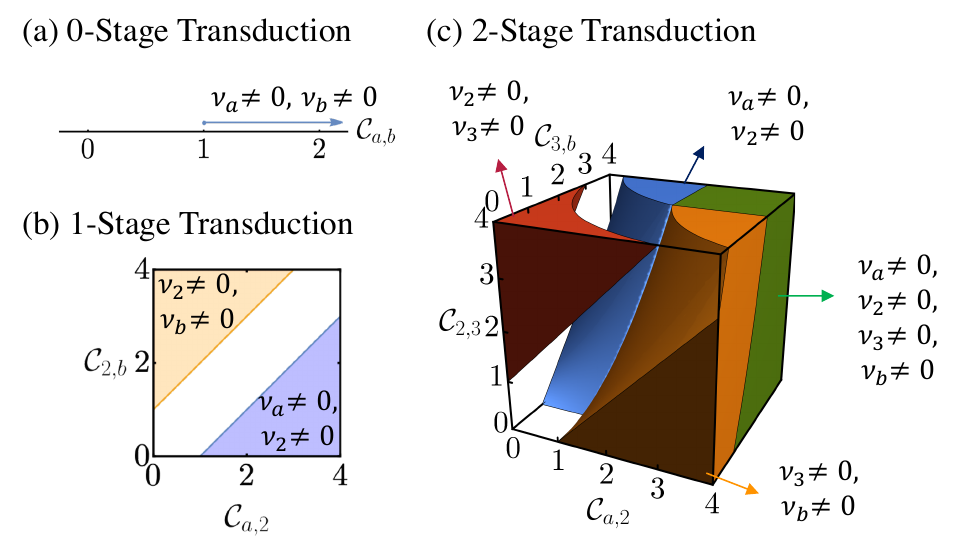}
\caption{Phase diagrams for 0-, 1-, and 2-stage transduction in the parameter space of mode cooperativities.  In the colored regimes, maximal internal efficiency is achieved with off-resonant frequencies as labeled.}
\label{fig:PhaseDiagram} 
\end{center}
\end{figure}

For 0-, 1-, and 2-stage transduction, we identify the regimes where maximal internal conversion efficiency is achieved with off-resonant frequencies $\nu_{j} \equiv \omega+ \D{j}\neq 0$ in the parameter space of mode cooperativities as shown in Fig.~\ref{fig:PhaseDiagram}~\cite{Note1}. These examples manifest the exotic behavior that a quantum transducer operating at off-resonant frequencies may outperform all-resonant ones.

Our method may be extended for conversion through an ensemble of intermediate modes.  For instance, considering transduction schemes mediated by two excited levels of a large number of atoms $N_A$~\cite{Everts2019,Bartholomew2020},  we can treat the atomic excited states as collective modes  $\hat{S}_{2(3)}=\frac{1}{\sqrt{N_A}}\sum_k \hat{m}_{2(3),k}$  and obtain the same form of conversion efficiency as in Eq.~(\ref{eqn:etaN}) but with enhanced coupling rates $\g{a(b)} \rightarrow \sqrt{N_A}\g{a(b)}$.  To include the effect of inhomogeneous broadening by taking the continuous limit $\sum_k \rightarrow N_A \int_{-\infty}^{\infty} d\omega_{2,k} d\omega_{3,k} \rho(\omega_{2,k}) \rho(\omega_{3,k})$
and assuming Lorentzian energy level distributions
$\rho(\omega_{2(3),k})= \frac{\Gamma_{2(3)}/2\pi}{(\omega_{2(3),k}-\omega_{2(3)})^2+(\Gamma_{2(3)}/2)^2}$, one can show that the updated conversion efficiency formula is associated with the broadened linewidth $\kappa_{2(3)} \rightarrow \kappa_{2(3)}+\Gamma_{2(3)}$. We may also expand our discussion to include thermal noise, which will introduce added noise to the transduction~\cite{Note1}.

\paragraph*{Conclusion and outlook.}
In conclusion, we have presented the generalized matching condition for $N$-stage quantum transduction and suggested different regimes of non-resonant conversions that can outperform all-resonant ones. Moreover, we drew a rigorous connection between transducer models and electric circuits, which brought the available circuit design toolboxes into this field. While our discussion has been focused on quantum transducers, it may also apply to general physical systems described by the externally-coupled bosonic-chain model for other applications including microwave photon detectors~\cite{Lescanne2020}, optical delay lines~\cite{Melloni2008}, and optical buffers~\cite{Xia2006}. Our work provides a generic formalism for determining experimental parameters to realize efficient quantum transduction in various platforms, with potential future extensions for transduction schemes utilizing two-mode-squeezing interactions~\cite{Zhong2020a,Zhong2022} or inhomogeneous couplings of the mediating ensemble.

\begin{acknowledgments}
{We thank Vinicius Ferreira, Oscar Painter, Hong Tang, and Changling Zou for helpful discussions. We acknowledge support from the ARO (W911NF-18-1-0020, W911NF-18-1-0212), ARO MURI (W911NF-16-1-0349, W911NF-21-1-0325), AFOSR MURI (FA9550-19-1-0399, FA9550-21-1-0209), AFRL (FA8649-21-P-0781), DoE Q-NEXT, NSF (OMA-1936118, EEC-1941583, OMA-2137642), NTT Research, and the Packard Foundation (2020-71479)}.
\end{acknowledgments}

\bibliographystyle{apsrev4-1}
\bibliography{MatchRef}

\end{document}


\begin{center}
\textbf{\large{Supplemental Information for  ''Generalized Matching Condition for Unity Efficiency Quantum Transduction"}}\\
Chiao-Hsuan Wang, Mengzhen Zhang, and Liang Jiang

\end{center}

\section{Conversion Efficiency Derivation}
We characterize the dynamics of $N$-stage transduction using the standard input-output theory~\cite{Walls2008} and derive the conversion efficiency.  The Heisenberg Langevin equations of motion for the system modes in the frequency domain are
\begin{align}
&\ch{a}^{-1}\ad[\omega]= i \g{1}\md{2}[\omega] +\sqrt{\kappa_{a, \rm ex}}\ad_{\rm in,ex}[\omega]+\sqrt{\kappa_{a, \rm i}}\hat{A}^{\dagger}_{\rm in}[\omega], \notag\\
&\ch{j}^{-1}\md{j}[\omega]= i\g{j} \md{j+1}[\omega] +i\g{j-1}\md{j-1}[\omega]+\sqrt{\ka{j}}\hat{M}^{\dagger}_{\rm j, in}[\omega], \notag\\
&\chi_{b}^{-1}\bd[\omega]= i g_{N+1}\hat{m}^{\dagger}_{N+1}[\omega] +\sqrt{\kappa_{b, \rm ex}}\bd_{\rm in,ex}[\omega]+\sqrt{\kappa_{b, \rm i}}\hat{B}^{\dagger}_{\rm in}[\omega],
\label{eqn:eom}
\end{align}
where $\kappa_{a(b)\rm ex}$ is the coupling rate to the external channel with an input signal field $\ad(\bd)_{\rm in, ex}[\omega]$, $\kappa_{a(b),\rm i}$ is the intrinsic loss rate of mode $\ad (\bd)$ due to the coupling with input noise $\hat{A}(\hat{B})^{\dagger}_{\rm in}[\omega]$ from  the dissipative environment, and $\ka{j}$ is the intrinsic loss rate of the intermediate mode $\md{j}$ due to the coupling with input noise $\hat{M}^{\dagger}_{\rm j, in}[\omega]$. The end mode $\ad (\bd)[\omega]$ is subject to a total dissipation rate $\kappa_{a(b)}=\kappa_{a(b),\rm ex}+\kappa_{a(b),\rm i}$, and $\chi_{j}$'s are the mode susceptibilities defined as $
\chi_{j}^{-1}\equiv i(\omega+\D{j})+\ka{j}/2$.
For notational simplicity, we will drop the $[\omega]$ symbol for all the frequency domain mode operators from now on.

The input and output fields are connected to the bosonic modes long the chain according to the input-output relations
\begin{align}
&\ad_{\rm out,ex}=\ad_{\rm in,ex}-\sqrt{\kappa_{a, \rm ex}}\ad,
\hat{A}^{\dagger}_{\rm out}=\hat{A}^{\dagger}_{\rm in}-\sqrt{\kappa_{a, \rm i}}\ad,\notag\\
&\hat{M}^{\dagger}_{\rm j,out}=\hat{M}^{\dagger}_{\rm j,in}-\sqrt{\ka{j}}\md{j},\notag\\
&\bd_{\rm out,ex}=\bd_{\rm in,ex}-\sqrt{\kappa_{b, \rm ex}}\bd, \hat{B}^{\dagger}_{\rm out}=\hat{B}^{\dagger}_{\rm in}-\sqrt{\kappa_{b, \rm i}}\bd.
\label{eqn:inout}
\end{align}

Consider an input mode $\ad_{\rm in,ex}$ only, we can iteratively solve the equations of motion Eq.~(\ref{eqn:eom}) along with the input-output relations Eq.~(\ref{eqn:inout}),
\begin{align}
&\ad= \ch{a} (i \g{1}\md{2} +\sqrt{\kappa_{a, \rm ex}}\ad_{\rm in,ex}), \notag\\
&\md{2}= \frac{i\g{2}\ch{a}^{-1} \md{3}[\omega] +i\g{a}\sqrt{\kappa_{a, \rm ex}}\ad_{\rm in,ex}}{\left(\ch{2}^{-1}+\cfrac{g_a^2} {\ch{a}^{-1}}\right)\ch{a}^{-1}}, \notag\\
&\vdots\notag\\
&\bd_{\rm out,ex}= -\frac{\sqrt{\kappa_{a, \rm ex} \kappa_{b, \rm ex}}\prod_{j=1}^{N+1}(i g_j)}{\prod_{j=1}^{N+2}(\chi^{-1}_{j,\rm eff,r})}\ad_{\rm in,ex}=-\frac{\sqrt{\kappa_{a, \rm ex} \kappa_{b, \rm ex}}\prod_{j=1}^{N+1}(i g_j)}{\prod_{j=1}^{N+2}(\chi^{-1}_{j,\rm eff})}\ad_{\rm in,ex},
\label{eqn:solving}
\end{align}
where $\chi_{j,\rm eff}$ is the effective mode susceptibility viewed from the left due to the couplings $g_{j}, \cdots, g_{b}$,
\begin{align}
\ch{j,\rm eff}^{-1} \equiv \ch{j}^{-1}+\cfrac{\g{j}^2}{\ch{j+1}^{-1}+\cfrac{\g{j+1}^2}{\ch{j+2}^{-1}+\cfrac{\ddots}{\ddots+\cfrac{\g{b}^2}{\ch{b}^{-1}}}}},
\label{eqn:chieff}
\end{align}
and $\chi_{j,\rm eff,r}$ is the effective mode susceptibility viewed from the reverse direction due to the couplings $g_{j-1}, \cdots, g_{a}$,
\begin{align}
\ch{j,\rm eff,r}^{-1} \equiv \ch{j}^{-1}+\cfrac{\g{j-1}^2}{\ch{j-1}^{-1}+\cfrac{\g{j-2}^2}{\ch{j-2}^{-1}+\cfrac{\ddots}{\ddots+\cfrac{\g{a}^2}{\ch{a}^{-1}}}}}.
\label{eqn:chieffr}
\end{align}
The photon(boson) number conversion efficiency is defined as
\begin{align}
\eta_N[\omega]=\frac{\left\langle \bd_{\rm out,ex} \bh_{\rm out, ex}\right\rangle}{\left \langle \ad_{\rm in,ex} \ah_{\rm in, ex}\right\rangle}=\abs{\frac{\sqrt{\kappa_{a, \rm ex} \kappa_{b, \rm ex}}\prod_{j=1}^{N+1}(i g_j)}{\prod_{j=1}^{N+2}\chi^{-1}_{j,\rm eff}}}^2=\frac{\left\langle \ad_{\rm out,ex} \ah_{\rm out, ex}\right\rangle}{\left \langle \bd_{\rm in,ex} \bh_{\rm in, ex}\right\rangle}.
\end{align}
The transduction process is reciprocal with identical conversion efficiency for $\ad \rightarrow \bd$ and for $\bd \rightarrow \ad$.

Alternatively, one can find a scattering matrix $\mathcal{S}_{N}[\omega]$ between the input and output modes such that $\vec{a}^{\dagger}_{\rm out}[\omega]=\mathcal{S}_{N}[\omega]\vec{a}^{\dagger}_{\rm in}[\omega]$, where $\vec{a}^{\dagger}_{\rm in}[\omega]=(\ad_{\rm in,ex},\hat{A}^{\dagger}_{\rm in},\hat{M}^{\dagger}_{\rm 2, in},\cdots,\hat{B}^{\dagger}_{\rm in},\bd_{\rm in,ex})^{T}$ and $\vec{a}^{\dagger}_{\rm out}[\omega]=(\ad_{\rm out,ex},\hat{A}^{\dagger}_{\rm out},\hat{M}^{\dagger}_{\rm 2,out},\cdots,\hat{B}^{\dagger}_{\rm out},\bd_{\rm out,ex})^{T}$. By directly solving for the matrix, the conversion efficiency is given by

\begin{align}
    \eta_N[\omega]&\equiv \abs{\mathcal{S}_{N,\bd_{\rm out,ex} \ad_{\rm in,ex}}[\omega]}^2=\abs{\frac{\sqrt{\kappa_{a, \rm ex} \kappa_{b, \rm ex}}\prod_{j=1}^{N+1}(i g_j)}{D_N[\omega]}}^2=\abs{\mathcal{S}_{N,\ad_{\rm out,ex} \bd_{\rm in,ex}}[\omega]}^2,
\end{align}
where $D_N[\omega]$ is the determinant of a ($N$+2)-dimensional tridiagonal matrix
\begin{align}
D_N[\omega] \equiv \begin{vmatrix}
\ch{a}^{-1} & i\g{a} & 0& \cdots & \cdots &0\\
 i\g{a} & \ch{2}^{-1} & ig_{2}  &\ddots & & \vdots\\
0 & i\g{2} & \ddots & \ddots  & \ddots  &  \vdots\\
\vdots & \ddots &  \ddots   & \ddots  & \ddots &  0 \\
\vdots &  & \ddots  & \ddots & \ddots &  i\g{b}\\
 0 & \cdots & \cdots  & 0 & i\g{b}& \ch{b}^{-1}
\end{vmatrix}.
\label{eqn:DN}
\end{align}
We can show that the two results agree with each other by using a mathematical theorem for tridiagonal matrices \cite{Kilic2008},
\begin{align}
\begin{vmatrix}
a_1 & b_1 & 0& \cdots & \cdots &0\\
c_1 & a_2 & b_2  &\ddots & & \vdots\\
0 & c_2 & \ddots & \ddots  & \ddots  &  \vdots\\
\vdots & \ddots &  \ddots   & \ddots  & \ddots &  0 \\
\vdots &  & \ddots  & \ddots & \ddots &  b_{n-1}\\
 0 & \cdots & \cdots  & 0 & c_{n-1}& a_n
\end{vmatrix}&=\begin{vmatrix}
a_1 & b_1 & 0& \cdots & \cdots &0\\
c_1 & a_2 & b_2  &\ddots & & \vdots\\
0 & c_2 & \ddots & \ddots  & \ddots  &  \vdots\\
\vdots & \ddots &  \ddots   & \ddots  & \ddots &  0 \\
\vdots &  & \ddots  & \ddots & \ddots &  b_{n-1}\\
 0 & \cdots & \cdots  & 0 & c_{n-1}& a_n
\end{vmatrix}\notag\\
=\prod_{j=1}^{n} a_j+\cfrac{- b_{j-1} c_{j-1}}{a_{j-1}+\cfrac{- b_{j-2} c_{j-2}}{a_{j-2}+\cfrac{\ddots}{\ddots+\cfrac{- b_1 c_1}{a_1}}}}
&=\prod_{j=1}^{n} a_j+\cfrac{- b_{j} c_{j}}{a_{j+1}+\cfrac{- b_{j+1} c_{j+1}}{a_{j+2}+\cfrac{\ddots}{\ddots+\cfrac{- b_{n-1}c_{n-1}}{a_n}}}}.
\label{eqn:tridiagonal}
\end{align}

\section{Effective Electric Circuit Model}
\begin{figure}[htbp]
\begin{center}
\includegraphics[width=0.8 \textwidth]{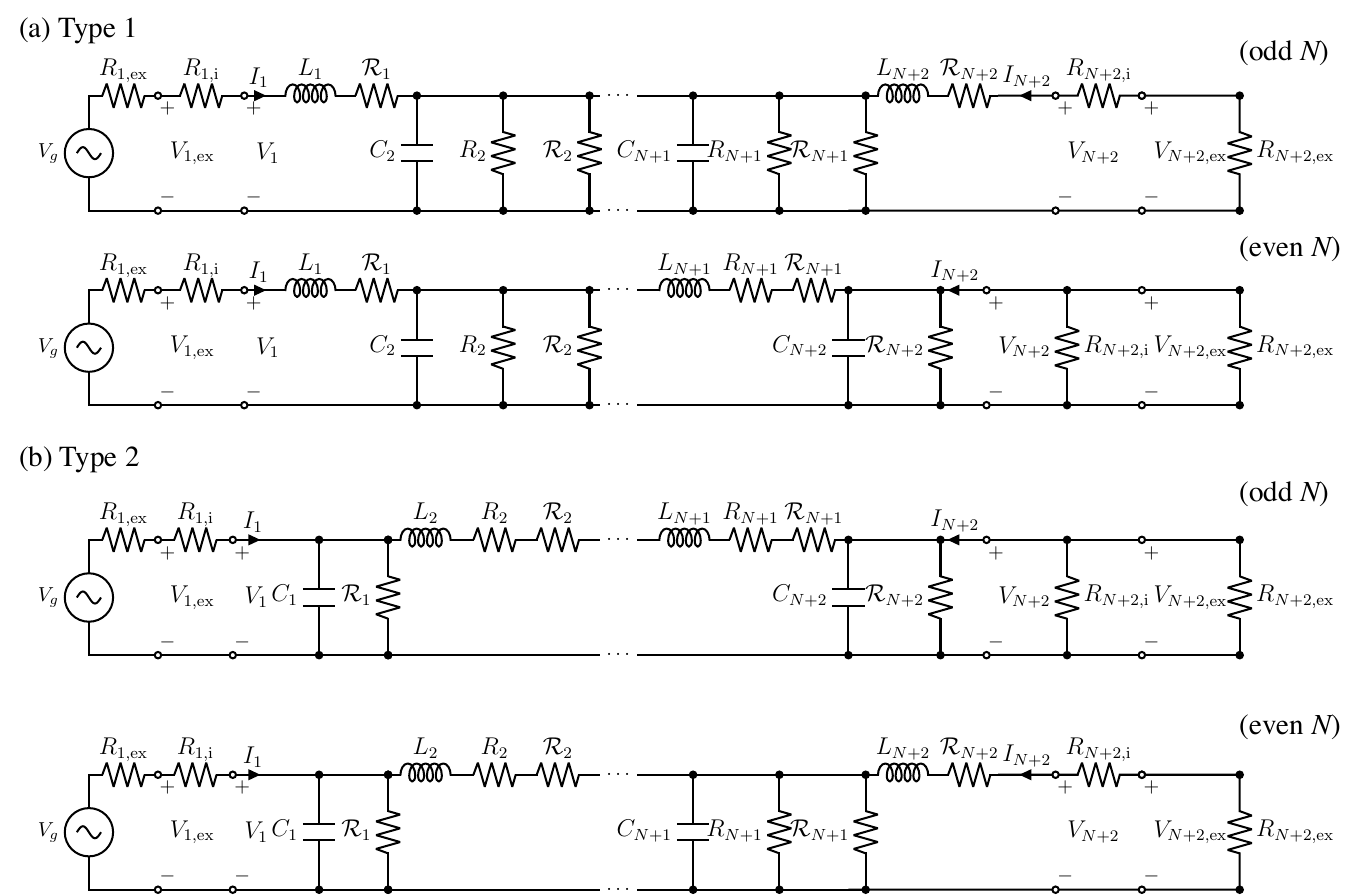}
\caption{Schematic of effective network circuits that give rise to identical frequency response as a $N$-stage quantum transducer. (a) Type 1 network circuit models that start with a series inductor.  (b) Type 2 network circuit models that start with a parallel capacitor.}
\label{fig:circuitNeff} 
\end{center}
\end{figure}

Consider a two-port network circuit in Fig.~\ref{fig:circuitNeff}(a) with inductance $L_{j}  \in \mathbb{R}$, capacitance $C_{j} \in \mathbb{R}$, resistance $R_{j}=1/G_{j} \in \mathbb{R}$, and generalized resistance $\mathcal{R}_{j}=1/\mathcal{G}_{j} \in i \mathbb{R}$.
The internal transmission coefficient for power waves can be calculated as~\cite{Youla1971,Kurokawa1965}
\begin{align}
t^p_{N,\rm int}[\omega]=\mathcal{S}^p_{N,\textrm{int }\beta_{\rm out} \alpha_{\rm in}}&=2\sqrt{\frac{R_{1}}{R_{N+2}}}\frac{V_{N+2}[\omega]}{V_{g}[\omega]}
=\begin{cases}
2\sqrt{R_{1} R_{N+2}} \prod_{j=1}^{N+2} \X{j,\rm eff}^{-1}& \text{odd }N\\
2\sqrt{R_{1}/R_{N+2}}\prod_{j=1}^{N+2} \X{j,\rm eff}^{-1} & \text{even }N
\end{cases}.
\label{eqn:TNint}
\end{align}
where $X_{j,\rm eff}$ is the effective impedance $Z_{j,\rm eff}$ (for odd $j$) or admittance $Y_{j,\rm eff}$(for even $j$) viewed from the $ j-1$ element,
\begin{align}
\X{j,\rm eff} \equiv X_{j}+\cfrac{1}{\X{j+1}+\cfrac{1}{\X{j+2}+\cfrac{\ddots}{\ddots +\X{N+2}}}},
\label{eqn:Zeff}
\end{align}
where $X_{j}$ is the bare impedance or admittance, $X_{j} \equiv i \omega L_{j} +R_{j}+ \mathcal{R}_{j}=Z_{j}$ for odd $j$ and $X_{j} \equiv i \omega C_{j} +G_{j}+ \mathcal{G}_{j}=Y_{j}$ for even $j$.
By setting
\begin{align}
\begin{cases}
L^{-1}_{j} C^{-1}_{j+1}=\g{j}^{2}, \,R_{j} L^{-1}_{j} =\ka{j}/2, \, \mathcal{R}_{j} L^{-1}_{j} = i \Delta_j  & \text{odd }j\\
C^{-1}_{j} L^{-1}_{j+1}=\g{j}^{2}, \, G_{j} C^{-1}_{j} =\ka{j}/2, \, \mathcal{G}_{j} C^{-1}_{j} = i \D{j} & \text{even }j
\end{cases},
\label{eqn:circuitparas}
\end{align}
one can show that $\X{j, \rm eff}/L_{j}=\ch{j,\rm eff}^{-1}$ for odd $\rm j$, $\X{j, \rm eff}/C_{j} =\ch{j,\rm eff}^{-1}$ for even $\rm j$, and that
\begin{align}
t^p_{N,\rm int}[\omega] =\frac{\sqrt{\kappa_{a} \kappa_{b}}\prod_{\rm j=1}^{N+1}\g{j}}{\prod_{\rm j=1}^{N+2}\ch{j,\rm eff}^{-1}}.
\label{eqn:Tint=Sint}
\end{align}
The effective circuit renders a transmission coefficient that coincides with the transducer scattering matrix element $\mathcal{S}_{N,\bd_{\rm out,ex} \ad_{\rm in,ex}}$ up to a prefactor $-(i)^{N+1} \sqrt{\eta_{\rm ext}}$.  The input and output modes are analogous to the incident and reflected power waves of the circuit. The explicit analogy between the transducer system and the effective electric circuit are summarized in table~\ref{table:analogy}.

We conclude that the internal power efficiency of the circuit is $\eta^p_{N,\rm int}[\omega] \equiv \abs{t^p_{N,\rm int}[\omega]}^2=\eta_{N,\rm int}[\omega]$.  One may also separate the external coupling rates from the total dissipation rates of the end modes to replicate the full conversion efficiency $\eta^p_{N}[\omega] \equiv \abs{t^p_N[\omega]}^2=\eta_{N}[\omega]$. Specifically,
\begin{align}
&R_{1,\rm ex} L^{-1}_{1} =\ka{a, \rm ex}/2, \,R_{1,\rm i} L^{-1}_{1} =\ka{a, \rm i}/2, \,R_{1}=R_{1,\rm ex}+R_{1,\rm i},\\
&\begin{cases}
R_{N+2,\rm ex} L^{-1}_{N+2} =\ka{b, \rm ex}/2, \,R_{N+2,\rm i} L^{-1}_{N+2} =\ka{b, \rm i}/2, \,R_{N+2}=R_{N+2,\rm ex}+R_{N+2,\rm i} &\text{odd }N\\
G_{N+2, \rm ex} C^{-1}_{N+2} =\ka{b, \rm ex}/2, \, G_{N+2, \rm i} C^{-1}_{N+2} =\ka{b, \rm i}/2, \,G_{N+2}=G_{N+2,\rm ex}+G_{N+2,\rm i} &\text{even }N
\end{cases},
\label{eqn:Rends}
\end{align}

\begin{align}
t^p_N[\omega]=S^p_{N,\beta_{\rm out,ex} \alpha_{\rm in, ex}}&=2\sqrt{\frac{R_{1,\rm ex}}{R_{N+2,\rm ex}}}\frac{V_{N+2,\rm ex}[\omega]}{V_{g}[\omega]}
=\begin{cases}
\sqrt{\frac{R_{1,\rm ex}}{R_{1}}\frac{R_{N+2,\rm ex}}{R_{N+2}}}t^p_{N,\rm int}[\omega]& \text{odd }N\\
\sqrt{\frac{R_{1,\rm ex}}{R_{1}}\frac{R_{N+2}}{R_{N+2,\rm ex}}}t^p_{N,\rm int}[\omega] & \text{even }N
\end{cases}.
\label{eqn:TN}
\end{align}
One can also verify that the effective circuits of type 2 form in Fig.~\ref{fig:circuitNeff}(b) possess identical transmission coefficients as type 1 circuits (Fig.~\ref{fig:circuitNeff}(a)) by interchanging the values of $L_{j}$ with $C_{j}$, $\mathcal{R}_{j}$ with $\mathcal{G}_{j}$, and $R_{j}$ with $G_{j}$.

\begin{table}
\caption{Analogy between the transducer system and the effective electric circuit}
\begin{ruledtabular}
\begin{tabular}{ l l l l l l l l }
Transducer System & Electric Circuit\\
$\ad_{\rm in,ex}$ & $c_s\footnote[1]{$c_s$ is some dimensional scaling factor.}\alpha_{\rm in,ex}=\frac{c_s V_g}{2 \sqrt{R_{1,\rm ex}}}$\\
$\bd_{\rm out,ex}$ &$c_s\beta_{\rm out,ex}=\frac{c_s V_{N+2,\rm ex}}{\sqrt{R_{N+2,\rm ex}}}$\\
$\ad_{\rm in}$ & $c_s\alpha_{\rm in}=\frac{c_s V_g}{2 \sqrt{R_{1}}}$ \\
$\bd_{\rm out}$ & $c_s\beta_{\rm out}=\frac{c_s V_{N+2}}{\sqrt{R_{N+2}}}$\\
$\hat{a}^{\dagger}$ & $\frac{c_s I_1 \sqrt{R_{1}}}{\sqrt{\kappa_{a}}}$\\
$\hat{b}^{\dagger}$ & $\frac{c_s I_{N+2} \sqrt{R_{N+2}}}{\sqrt{\kappa_{b}}}$\\ \hline
\end{tabular}
\begin{tabular}{ l l l l l l l l }
Transducer System & Electric Circuit (odd $j$) & Electric Circuit (even $j$) \\
$\g{j}$ & $\sqrt{L^{-1}_{j} C^{-1}_{j+1}}$ & $\sqrt{C^{-1}_{j} L^{-1}_{j+1}}$\\
$\ka{j}/2$ & $R_{j} L^{-1}_{j}$ & $G_{j} C^{-1}_{j}$\\
$i \Delta_j$ & $\mathcal{R}_{j} L^{-1}_{j}$ & $\mathcal{G}_{j} C^{-1}_{j}$\\ 
$\chi_{j}^{-1}$ & $Z_{j}/L_{j}$  & $Y_{j}/C_{j}$\\
$\chi^{-1}_{j,\rm eff}$ & $Z_{j,\rm eff}/L_{j}$& $Y_{j,\rm eff}/C_{j}$\\
$\chi^{-1}_{j,\rm eff}-\chi^{-1}_{j}=\g{j}^2 \chi_{j+1,\rm eff}$ & $(Z_{j,\rm eff}-Z_{j})/L_{j}=Y_{j+1,\rm eff}^{-1}/L_{j}$  & $(Y_{j,\rm eff}-Y_{j})/C_{j}=Z_{j+1,\rm eff}^{-1}/C_{j}$\\ \hline
\end{tabular}
\end{ruledtabular}
\label{table:analogy}
\end{table}

\section{Generalized Matching Condition}
The matching condition viewed at the $\ad$ mode is
\begin{align}
(\ch{a}^{-1})^{*}=\g{a}^2 \chi_{2,\rm eff}.
\label{eqn:amatchA}
\end{align}
Reorganizing the condition into $-(\ch{a}^{-1})^{*}+\g{a}^2 \chi_{2,\rm eff}=0$ and multiply the expressions by $\prod_{j=2}^{N+2} \chi_{j,\rm eff}^{-1}$, where all $\chi_{j,\rm eff}${'}s are all nonzero with $\kappa_{a},\kappa_{b}, g_j\text{{'}s} \neq 0$, we can again use the theorem for tridiagonal matrices \ref{eqn:tridiagonal} to show that
\begin{align}
(-(\ch{a}^{-1})^{*}+\g{a}^2 \chi_{2,\rm eff})\prod_{j=2}^{N+2} \chi_{j,\rm eff}^{-1} = \begin{vmatrix}
-(\ch{a}^{-1})^{*} & i\g{a} & 0& \cdots & \cdots &0\\
 i\g{a} & \ch{2}^{-1} & i g_2  &\ddots & & \vdots\\
0 & i\g{2} & \ddots & \ddots  & \ddots  &  \vdots\\
\vdots & \ddots &  \ddots   & \ddots  & \ddots &  0 \\
\vdots &  & \ddots  & \ddots & \ddots &  i g_{b}\\
 0 & \cdots & \cdots  & 0 & i g_{b}&\ch{b}^{-1}
\end{vmatrix}.
\label{eqn:MNA1}
\end{align}
Combining with the condition of no intermediate loss, we arrive at the generalized matching condition in a matrix form
\begin{align}
M_{N}\equiv \left.\begin{vmatrix}
-(\ch{a}^{-1})^{*} & i\g{a} & 0& \cdots & \cdots &0\\
 i\g{a} & \ch{2}^{-1} & i g_2  &\ddots & & \vdots\\
0 & i\g{2} & \ddots & \ddots  & \ddots  &  \vdots\\
\vdots & \ddots &  \ddots   & \ddots  & \ddots &  0 \\
\vdots &  & \ddots  & \ddots & \ddots &  i g_{b}\\
 0 & \cdots & \cdots  & 0 & i g_{b}&\ch{b}^{-1}
\end{vmatrix} \right|_{\kappa_{2}=\cdots=\kappa_{N+1}=0}=0.
\label{eqn:MNA}
\end{align}

When the generalized matching condition is satisfied, the added noise for the conversion can be expressed in a simple form. The added noise for conversion from $\ad$ to $\bd$, relative to the signal, is quantified as
\begin{align}
\left\langle \ad_{\rm out,ex} \ah_{\rm out,ex}\right\rangle=\eta_{N}\left(\left\langle\bd_{\rm in,ex}\bh_{\rm in,ex}\right\rangle+n^{ a\rightarrow b}_{\rm add} \right), 
\label{eqn:noise}
\end{align}
where the angular bracket represents the quantum statistical average over a density matrix. For the impedance-matched $N$-stage tranduction with no intermediate loss, the added noise from $\ad$ to $\bd$ is
\begin{align}
n^{a \rightarrow b}_{\rm add}[\omega]=\frac{\kappa_{a, \rm i}}{\kappa_{a, \rm ex}} \left\langle \hat{B}^{\dagger}_{\rm in}\hat{B}_{\rm in}\right\rangle + \frac{\kappa_{a}}{\kappa_{a, \rm ex}} \frac{\kappa_{b, \rm i}}{\kappa_{b}}\left\langle \hat{A}^{\dagger}_{\rm in}\hat{A}_{\rm in}\right\rangle + \frac{\kappa_{a}}{\kappa_{a, \rm ex}} \frac{\kappa_{b}}{\kappa_{b, \rm ex}} \frac{\kappa_{b, \rm i}^2}{\kappa_{b}^2} \left\langle \hat{a}^{\dagger}_{\rm in, ex}\hat{a}_{\rm in, ex}\right\rangle.
\label{eqn:naddNab}
\end{align}
The added noise for the impedance-matched $N$-stage transduction from $\bd$ to $\ad$ can be found analogously,
\begin{align}
n^{b \rightarrow a}_{\rm add}[\omega]=\frac{\kappa_{b, \rm i}}{\kappa_{b, \rm ex}} \left\langle \hat{A}^{\dagger}_{\rm in}\hat{A}_{\rm in}\right\rangle + \frac{\kappa_{b}}{\kappa_{b, \rm ex}} \frac{\kappa_{a, \rm i}}{\kappa_{a}}\left\langle \hat{B}^{\dagger}_{\rm in}\hat{B}_{\rm in}\right\rangle + \frac{\kappa_{b}}{\kappa_{b, \rm ex}} \frac{\kappa_{a}}{\kappa_{a, \rm ex}} \frac{\kappa_{a, \rm i}^2}{\kappa_{a}^2} \left\langle \hat{b}^{\dagger}_{\rm in, ex}\hat{b}_{\rm in, ex}\right\rangle.
\label{eqn:naddNbab}
\end{align}
Large added noise from the thermal environment will decrease the fidelity of the transferred quantum state.

\section{Impedance Matching for lossy 1-Stage Transduction}

\begin{figure}[tbp]
\begin{center}
\includegraphics[width=0.7 \textwidth]{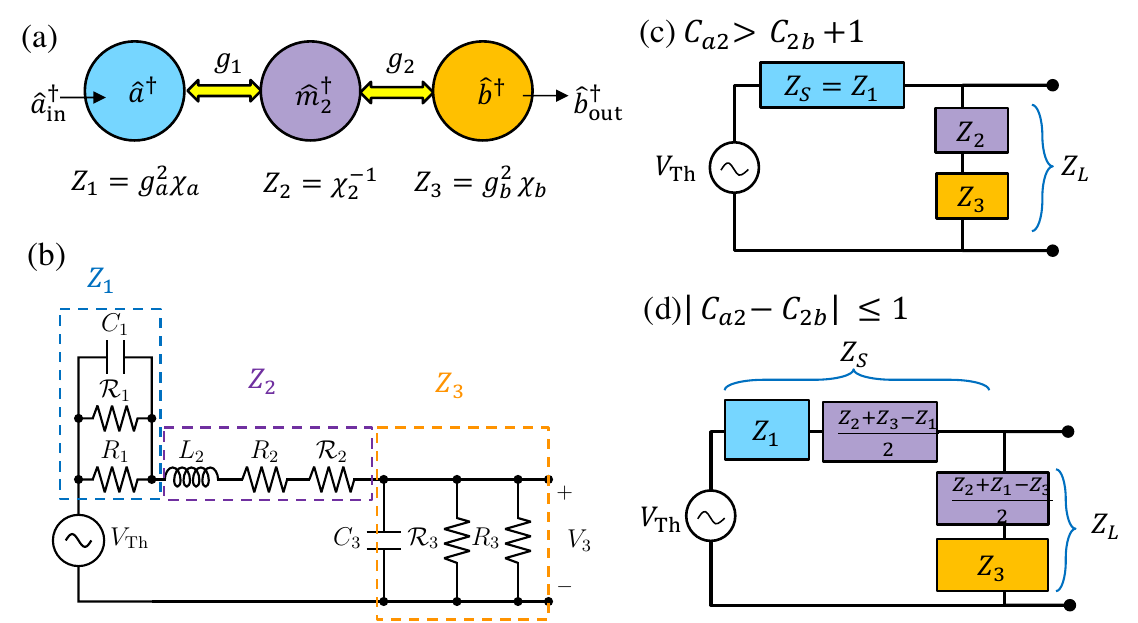}
\caption{(a) Schematic of 1-stage quantum transduction system. (b) Effective circuit that reproduces the 1-stage internal conversion efficiency by setting $C_1=1/\g{a}^{2}$, $\mathcal{R}_1=\g{a}^2/(i\D{a})$, $R_1=2\g{a}^2/\kappa_{a}$, $L_2=1$, $R_2=\ka{2}/2$, $\mathcal{R}_2=i \D{2}$, $C_3=1/\g{b}^{2}$, $\mathcal{R}_3=\g{b}^2/(i\D{b})$, $R_3=2\g{b}^2/\kappa_{b}$, and $V_{\rm Th}=V_{g}/R_1/( i\omega C_1  +\mathcal{G}_1+ G_1)$. This is the Thevenin equivalent circuit of the type 2 effective circuit mode for 1-stge transduction. (c) Impedance matched lossy circuit configuration if $\mathcal{C}_{a2} > \mathcal{C}_{2b}+1$.  (d) Impedance matched lossy circuit configuration if $\abs{\mathcal{C}_{a2}- \mathcal{C}_{2b}} \leq 1$.}
\label{fig:1stage} 
\end{center}
\end{figure}

In this section we discuss the detailed impedance matching interpretation for the optimal parameters of lossy 1-stage conversion (Fig.~\ref{fig:1stage}(a)). An effective circuit mode for 1-stage transduction is presented in Fig.~\ref{fig:1stage}(b). Here we consider at the Thevenin equivalent circuit for the type 2 form in Fig.~\ref{fig:circuitNeff}(b) and take the perspective from mode $\md{2}$ by taking $L_2=1$ such that the circuit interpretation will be symmetric in $\ad$ and $\bd$. In particular, with this choice of circuit, the middle mode $\md{2}$ has an impedance $Z_2=\chi_2^{-1}$, and the first mode $\ad$ has an effective impedance $Z_{1}=g_a^2 \chi_a$ while the third mode $\bd$ has $Z_{3}=g_b^2 \chi_b$.

When one of the mode is over-coupled, say if $\mathcal{C}_{a,2} > \mathcal{C}_{2,b}+1$, the maximal internal efficiency is achieved by the choice of optimal frequencies satisfying $\g{a}^2 \ch{a}^{*}=\ch{2}^{-1}+\g{b}^2 \ch{b}$ and $\nu_b=\omega+\D{b}=0$.
This condition can be understood in analogy to an impedance matched circuit with internal loss in Fig.~\ref{fig:1stage}(c). Treating the middle mode and the third mode altogether as the load, the system is impedance matched when $\g{a}^2 \ch{a}^{*}=Z_1^{*}=Z_2+Z_3=\chi_2^{-1}+g_b^2\chi_b$, and the maximal internal efficiency is $\eta_{\rm int}^{\rm max}=\mathcal{C}_{2,b}/(\mathcal{C}_{2,b}+1)$ when $\nu_{b}=0$, limited by the smaller cooperativity.  In other word, we are adding nonzero $\nu_{a}$ and $\nu_{2}$ to modify the source impedance such that it matches with the load impedance, which effectively reduce the source resistance $\Re[Z_1]$, while keeping the third mode at resonance, which gives rise to the maximal value of $\Re[Z_3]$, to get the largest fraction of power output at the load. Note that the internal reflection coefficient vanishes at the over-coupled port, $r_{1,a}=0$, while $r_{1,b} \neq 0$ due to the intermediate loss.

If $\abs{\mathcal{C}_{a,2} - \mathcal{C}_{2,b}}\leq 1$, the conversion is optimized when the system is all-resonant, $\nu_{a}=\nu_{ 2}=\nu_{b}=0$.
This condition can again be understood as an impedance matched lossy circuit (Fig.~\ref{fig:1stage}(d)). Treating the middle mode as a lossy component partly in the source and partly in the load, the impedance is matched at a point where $Z_{1}+Z_{S,2}=Z_{L,2}+Z_{3}$, which corresponds to $\mathcal{C}_{a,2} + (\mathcal{C}_{2,b}- \mathcal{C}_{a,2})/2= (\mathcal{C}_{a,2}- \mathcal{C}_{2,b})/2+\mathcal{C}_{2,b}$.  The maximal achievable internal conversion efficiency in this parameter regime is $\eta_{\rm int}^{\rm max}=4\mathcal{C}_{a,2}\mathcal{C}_{2,b}/(\mathcal{C}_{a,2}+\mathcal{C}_{2,b}+1)^2$.

\section{Optimal frequencies}
In this section we summarize the optimal frequencies for 1-stage and 2-stage transduction that give rise to maximal achievable internal efficiency at given cooperativities $\mathcal{C}_{j, j+1}=\frac{4g_j^2}{\kappa_j \kappa_{j+1}}$.

\begin{table}[htbp]
\caption{Optimal frequencies for 1-stage transduction}
\begin{ruledtabular}
\begin{tabular}{ l l l l l l l }
$N$=1  &$\kappa_2 =0$ & $\mathcal{C}_{a,2}+1 < \mathcal{C}_{2,b}$ & $\abs{\mathcal{C}_{a,2}- \mathcal{C}_{2,b}} \leq 1$ & $\mathcal{C}_{a,2} > \mathcal{C}_{2,b}$+1\\
 \hline
$\nu_{a}$  & \multirow{3}{2.5cm}{Multiple solutions satisfying $M_{1}=0$} &  0   &  0& $\pm \frac{\kappa_{a}}{2}\sqrt{\frac{\mathcal{C}_{a,2}}{\mathcal{C}_{2,b}+1}-1}$\\
$\nu_{\rm 2}$ & &  $\pm \frac{\kappa_2}{2}(\mathcal{C}_{a,2}+1)\sqrt{\frac{\mathcal{C}_{2,b}}{\mathcal{C}_{a,2}+1}-1}$   & 0 & $\pm \frac{\kappa_2}{2} ( \mathcal{C}_{2,b}+1)\sqrt{\frac{\mathcal{C}_{a,2}}{\mathcal{C}_{2,b}+1}-1}$\\
$\nu_{b}$&  & $\pm \frac{\kappa_{b}}{2}\sqrt{\frac{\mathcal{C}_{2,b}}{\mathcal{C}_{a,2}+1}-1}$   & 0 & 0
\\ $\eta_{\rm 1,int}^{\rm max}$ & 1
 & $\frac{\mathcal{C}_{a,2}}{\mathcal{C}_{a,2}+1}$ & $\frac{4 \mathcal{C}_{a,2} \mathcal{C}_{2,b}}{(\mathcal{C}_{a,2}+\mathcal{C}_{2,b}+1)^2}$ & $\frac{\mathcal{C}_{2,b}}{\mathcal{C}_{2,b}+1}$
\end{tabular}
\end{ruledtabular}
\end{table}

\begin{table}[htbp]
\caption{Optimal frequencies for 2-stage transduction}
\begin{ruledtabular}
\begin{tabular}{ l l l l l l l l l }
$N$=2  & \vbox{\hbox{\strut$\kappa_2=\kappa_3$}\hbox{\strut $\,\,\,\,\,\,\,\,=0$}} & \vbox{\hbox{\strut$\kappa_2=0,$} \hbox{\strut$\kappa_3 \neq 0$}}  & \vbox{\hbox{\strut$\kappa_2 \neq 0,$} \hbox{\strut$\kappa_3 = 0$}} & $\mathcal{C}_{2,3} > (\mathcal{C}_{a,2}+1)(\mathcal{C}_{3,b}+1)$ &   \vbox{\hbox{\strut $\abs{\mathcal{C}_{a,2}- \frac{\mathcal{C}_{2,3}}{\mathcal{C}_{3,b}+1}} \leq 1$} \hbox{\strut and $\abs{\mathcal{C}_{3,b}- \frac{\mathcal{C}_{2,3}}{\mathcal{C}_{a,2}+1}} \leq 1$}} \\ \\ \hline
$\nu_{a}$  & \multirow{4}{2cm}{Solutions satisfying $M_{2}=0$} & \multirow{3}{2cm}{Solutions satisfying $r_{2,a}=0$} &0 &  0   &  0\\
$\nu_{\rm 2}$ & & & \multirow{3}{2cm}{Solutions satisfying $r_{2,b}=0$} & $\pm \frac{\kappa_2(\mathcal{C}_{a,2}+1)\sqrt{\frac{\mathcal{C}_{2,3}}{(\mathcal{C}_{a,2}+1)(\mathcal{C}_{3,b}+1)}-1}}{2}$   & 0 \\
$\nu_{\rm 3}$ & & & &$\pm \frac{\kappa_3(\mathcal{C}_{3,b}+1)\sqrt{\frac{\mathcal{C}_{2,3}}{(\mathcal{C}_{a,2}+1)(\mathcal{C}_{3,b}+1)}-1}}{2}$  & 0\\
$\nu_{b}$&  & 0 & & 0 &  0 
\\ $\eta_{\rm 2,int}^{\rm max}$ & 1
 & $\frac{\mathcal{C}_{3,b}}{\mathcal{C}_{3,b}+1}$ & $\frac{\mathcal{C}_{a,2}}{\mathcal{C}_{a,2}+1}$ & $\frac{\mathcal{C}_{a,2}}{\mathcal{C}_{a,2}+1}\frac{\mathcal{C}_{3,b}}{\mathcal{C}_{3,b}+1}$ & $\frac{4 \mathcal{C}_{a,2} \mathcal{C}_{2,3} \mathcal{C}_{3,b}}{(\mathcal{C}_{a,2}+\mathcal{C}_{2,3}+\mathcal{C}_{3,b}+\mathcal{C}_{a,2}\mathcal{C}_{3,b}+1)^2}$
\end{tabular}
\end{ruledtabular}
\begin{ruledtabular}
\begin{tabular}{ l l l l l l l l l }
$N$=2  & \multicolumn{2}{c}{$\mathcal{C}_{a,2} \geq \mathcal{C}_{3,b}$ and $\mathcal{C}_{a,2} > \frac{\mathcal{C}_{2,3}}{\mathcal{C}_{3,b}+1}+1$} & \multicolumn{2}{c}{$\mathcal{C}_{3,b} \geq \mathcal{C}_{a,2}$ and $\mathcal{C}_{3,b} > \frac{\mathcal{C}_{2,3}}{\mathcal{C}_{a,2}+1}+1$}\\
& $\mathcal{C}_{3,b}^2 > \mathcal{C}_{2,3}+1$ &$\mathcal{C}_{3,b}^2 \leq \mathcal{C}_{2,3}+1$ & $\mathcal{C}_{a,2}^2 > \mathcal{C}_{2,3}+1$& $\mathcal{C}_{a,2}^2 \leq \mathcal{C}_{2,3}+1$
\\\hline
$\nu_{a}$  & $\pm \frac{\kappa_e}{2}\sqrt{\frac{\mathcal{C}_{a,2}}{\sqrt{\mathcal{C}_{2,3}+1}}-1}$ & $\pm \frac{\kappa_e}{2}\sqrt{\frac{\mathcal{C}_{a,2}}{\frac{\mathcal{C}_{2,3}}{\mathcal{C}_{3,b}+1}+1}-1}$ & $\pm \frac{\kappa_e}{2}\sqrt{\frac{\mathcal{C}_{a,2}}{\sqrt{\mathcal{C}_{2,3}+1}}-1}$ & 0\\
$\nu_{\rm 2}$ & $ \frac{\kappa_2}{\kappa_{a}}\sqrt{\mathcal{C}_{2,3}+1}\nu_{a}$ & $\frac{\kappa_2}{\kappa_{a}}\left(\frac{\mathcal{C}_{2,3}}{\mathcal{C}_{3,b}+1}+1 \right)\nu_{a}$ &$\frac{\kappa_2}{\kappa_{a}}\sqrt{\mathcal{C}_{2,3}+1}\nu_{a}$ &0\\
$\nu_{\rm 3}$ &$\frac{\kappa_3}{\kappa_{b}}\sqrt{\mathcal{C}_{2,3}+1}\nu_{b}$ & 0 &$\frac{\kappa_3}{\kappa_{b}}\sqrt{\mathcal{C}_{2,3}+1}\nu_{b}$ & $ \frac{\kappa_3}{\kappa_{b}}\left(\frac{\mathcal{C}_{2,3}}{\mathcal{C}_{a,2}+1}+1 \right)\nu_{b}$ \\
$\nu_{b}$& $\pm \mp \frac{\kappa_b}{2}\sqrt{\frac{\mathcal{C}_{3,b}}{\sqrt{\mathcal{C}_{2,3}+1}}-1}$ & 0 & $\pm \mp \frac{\kappa_b}{2}\sqrt{\frac{\mathcal{C}_{3,b}}{\sqrt{\mathcal{C}_{2,3}+1}}-1}$ & $\pm \frac{\kappa_b}{2}\sqrt{\frac{\mathcal{C}_{3,b}}{\frac{\mathcal{C}_{2,3}}{\mathcal{C}_{a,2}+1}+1}-1}$
\\ $\eta_{\rm 2,int}^{\rm max}$ & $\frac{\mathcal{C}_{2,3}}{(\sqrt{\mathcal{C}_{2,3}+1}+1)^2}$ & $\frac{\mathcal{C}_{,3b}}{\mathcal{C}_{3,b}+1}\frac{\mathcal{C}_{2,3}}{\mathcal{C}_{2,3}+\mathcal{C}_{3,b}+1}$ & $\frac{\mathcal{C}_{2,3}}{(\sqrt{\mathcal{C}_{2,3}+1}+1)^2}$ & $\frac{\mathcal{C}_{a,2}}{\mathcal{C}_{a,2}+1}\frac{\mathcal{C}_{2,3}}{\mathcal{C}_{a,2}+\mathcal{C}_{2,3}+1}$
\end{tabular}
\end{ruledtabular}
\end{table}

\bibliographystyle{apsrev4-1}
\bibliography{MatchRef}